# Highly coherent grain boundaries induced by local pseudo-mirror symmetry in $\beta$-Ga$_2$O$_3$

Yuchao Yan,[1,2] Yingying Liu,[3] Ziyi Wang,[1,2] Da Liu,[1,2] Xu Gao,[1,2] Yan Wang,[3] Cheng Li,[3] KeKe Ma,[3] Ning Xia,[3] Zhu Jin,[1,2,*] Tianqi Deng,[1,2,†] Hui Zhang,[1,2,‡] and Deren Yang[1,2]

[1]State Key Laboratory of Silicon and Advanced Semiconductor Materials and School of Materials Science and Engineering, Zhejiang University, Hangzhou 310027, China
[2]Institute of Advanced Semiconductors & Zhejiang Provincial Key Laboratory of Power Semiconductor Materials and Devices, Hangzhou Global Scientific and Technological Innovation Center, Zhejiang University, Hangzhou, Zhejiang 310027, China
[3]Hangzhou Garen Semiconductor Company Limited, Hangzhou, Zhejiang 311200, China

**ABSTRACT**. Grain boundaries have extensive influence on the performance of crystal materials. However, the atomic-scale structure and its relation with local and crystallographic symmetries remain elusive in low-symmetry crystals. Herein, we find that the local pseudo-mirror-symmetric atomic layer is the common physical origin of a series of highly coherent grain boundaries in the low-symmetry $\beta$-Ga$_2$O$_3$ crystal. These include the (100) twin boundary and an emerging series of $(h − 1'0'2)/(h + 1'0\overline{2})$ coherent asymmetric grain boundaries (CAGBs). Owing to the local pseudo-mirror symmetry and the special geometric relation of the $\beta$-Ga$_2$O$_3$ conventional cell, these CAGBs place 80% of the boundary atoms in pseudo-coincident sites, exhibiting high coherence under the coincident-site lattice model. With a combination of density functional theory calculations, Czochralski growth experiment, and atomic-scale characterizations, the structure and stability of the $(002)/(20\overline{2}) − A$ CAGB are confirmed, with a boundary energy density as low as 0.36 J/m$^2$. This CAGB is responsible for the spontaneous formation of a twinned defect facet at the surface steps during the epitaxy growth of $\beta$-Ga$_2$O$_3$, warranting a substrate orientation selection rule for $\beta$-Ga$_2$O$_3$. Through this study, we provide insights into the grain boundary physics in the low-symmetry $\beta$-Ga$_2$O$_3$ crystal while emphasizing the importance of the local pseudo-symmetries in the low-symmetry crystals.

$\beta$-Ga$_2$O$_3$ is an emerging ultra-wide-bandgap semiconductor with great potential in high-performance electronic and optical devices due to its outstanding physical properties and the feasibility of obtaining low-cost and high-quality single crystal substrates through molten-growth techniques [1-3]. However, the unexpected presence of twin boundaries during both bulk crystal and epitaxy growth profoundly affects the yield of high-quality substrates and high-performance devices, and hinders the commercialization process [4-8]. Recent experimental and first-principles studies on the twin boundaries in $\beta$-Ga$_2$O$_3$ have revealed the ultralow formation energy density of the (100) twin boundary [9-12]. Nevertheless, the complexity of grain boundaries arising from the low-symmetry C2/m space group of $\beta$-Ga$_2$O$_3$ has not been fully understood. To better understand the grain boundary physics and improve the quality of both bulk and epitaxy growth, atomic-scale investigations of the grain boundaries in $\beta$-Ga$_2$O$_3$ and their physical origins are urgently required.

In this Letter, the highly coherent (100) twin boundary and a series of coherent asymmetric grain boundaries (CAGBs) in the low-symmetry $\beta$-Ga$_2$O$_3$ crystal are investigated at the atomic-scale. They share a common microscopic origin, i.e., the local pseudo-mirror-symmetric atomic layers, as we will detail in this manuscript. Assisted by the local pseudo-mirror symmetry and the special geometric relation of the $\beta$-Ga$_2$O$_3$ conventional cell, these boundaries exhibit high coherence under the coincident-site lattice (CSL) model, with 80% of the boundary atoms placed in pseudo-coincident sites. Density functional theory (DFT) calculations are performed to elucidate the atom configurations and corresponding energy densities of the CAGBs. The $(002)/(20\overline{2}) − A$ CAGB is revealed to have the lowest normalized boundary energy density $\sigma_n$ of 0.36 J/m$^2$. Combined with the macroscopic geometry and atomic-scale observation of the CAGB in Czochralski-grown crystal, the stability of the $(002)/(20\overline{2}) − A$ CAGB is confirmed. This CAGB plays a critical role during the epitaxy growth of $\beta$-Ga$_2$O$_3$ by facilitating the spontaneous formation of a twinned defect facet at the surface steps, and a selection rule of $\beta$-Ga$_2$O$_3$ substrate orientation is proposed to avoid its formation. Based on this study, we provide perspectives into the grain boundary physics of the low-symmetry $\beta$-Ga$_2$O$_3$ crystal, while emphasizing the importance of the local pseudo-symmetries in the low-symmetry crystals.

We start by observing the local pseudo-mirror-symmetric atomic layers extending along the [001] orientation of the low-symmetry $\beta$-Ga$_2$O$_3$ crystal, as shown in Fig. 1(a). Considering the periodic conditions of the $C2/m$ space group of $\beta$-Ga$_2$O$_3$, the pseudo-mirror planes of these layers are perpendicular to the (100) and (010) planes, and the spacing between the pseudo-mirror planes is $\frac{1}{2}|[\mathbf{001}]|$. The mismatch between this local symmetry and the macroscopic symmetry of $\beta$-Ga$_2$O$_3$ produces the (100) twin boundary shown in Fig. 1(b). The atomic displacements and bond distortions at the (100) twin interface are greatly suppressed by the local pseudo-mirror symmetry, resulting in an ultralow 0.008 J/m$^2$ formation energy density of the (100) twin boundary. The methods are present in Supplemental Material 1 [18]. Notably, although the (100) twin boundary can be formed by mirroring across two distinct (100)-A and (100)-B planes, they are unified to the same structure due to the local pseudo-mirror symmetry as detailed in Supplemental Material 2 [18].

*Contact author: jinzhuu@zju.edu.cn
†Contact author: dengtq@zju.edu.cn
‡Contact author: msezhanghui@zju.edu.cn

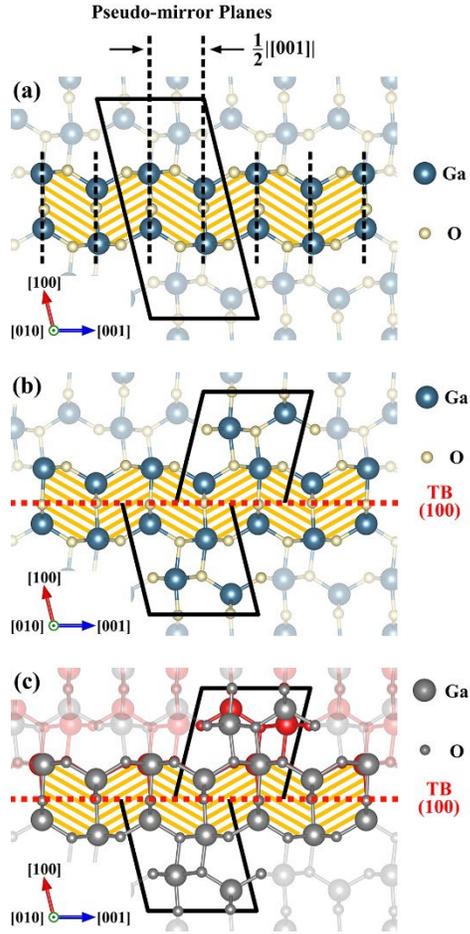

FIG. 1. (a) Schematic illustration of the local pseudo-mirror-symmetric atomic layer in β-Ga$_2$O$_3$. The pseudo-mirror planes are perpendicular to the (100) and (010) planes and marked by the dashed lines. (b) The (100) twin boundary based on the local pseudo-mirror-symmetric atomic layer, whose formation energy density is as low as 0.008 J/m$^2$. (c) The comparison between atoms around (100) twin boundary (red) and original crystal (gray) to show their pseudo-coincident relation.

The (100) twin boundary exhibits a high coherence under the coincident-site lattice (CSL) model by slightly relaxing the criterion of coincidence according to the pseudo-symmetry [13,14], i.e., the original and twinned atom sites are considered as pseudo-coincident if the displacement between them is less than 0.3 Å. Out of the ten atoms in half a $\beta$-Ga$_2$O$_3$ conventional cell (4 Ga atoms and 6 O atoms), only 2 Ga atoms are mismatched by a distance of 1.8 Å, as shown in Fig. 1(c). Consequently, 80 % of the atoms are in coincident sites, resulting in a reciprocal density $\Sigma$ of 1.25. This high coherence can readily explain the ultralow formation energy density of the (100) twin boundary and can be attributed to both the local pseudo-mirror-symmetric atomic layer shown in Fig. 1(a) and its in-plane offset shown in Fig. 2(a). The lattice parameters of $\beta$-Ga$_2$O$_3$ are a = 12.23 Å, b = 3.04 Å, c = 5.80 Å, and β = 103.7°, yielding a special lattice relationship of $|[\mathbf{100}]| \times \cos(\beta) \approx -\frac{1}{2}|[\mathbf{001}]|$ in Fig. 2(a). Hence, the in-plane offset between the adjacent pseudo-mirror-symmetric atomic layers is approximately $\frac{1}{4}|[\mathbf{001}]|$ along the [**001**] direction. This offset is only half of $\frac{1}{2}|[\mathbf{001}]|$, the spacing between the pseudo-mirror planes, and therefore the pseudo-mirror-symmetric atomic layer is mismatched with its two adjacent layers. However, two distinctive long-range, pseudo-coherent atom arrays that are almost perpendicular to the (100) plane can still be observed within one $\beta$-Ga$_2$O$_3$ conventional cell, as marked by the dashed line in Fig. 2(a). Consequently, the 6 oxygen atom sites in half a $\beta$-Ga$_2$O$_3$ conventional cell are all pseudo-coincident in the (100) twin boundary, while 2 Ga sites are pseudo-coincident and only the rest 2 Ga sites are mismatched, resulting in a very low formation energy density.

Apart from the (100) twin boundary, two coherent locally mirrored twin core structures can also be constructed by mirroring across one of the pseudo-coherent atom arrays, as shown in Fig. 2(b) and 2(c). These twin cores are geometrically consistent with the (100) twin boundary, i.e., their mirror planes are perpendicular to each other, suggesting that the (100) twin boundary can possibly be terminated and redirected into these twin cores. Furthermore, these twin cores exhibit the same coherence as the (100) twin boundary under the CSL model, suggesting their potentially low formation energy densities.

Considering the geometric and periodic conditions of the $\beta$-Ga$_2$O$_3$ conventional cell, the periodic stacking of these two local twin cores with different in-plane offsets $\frac{h}{2}[\mathbf{001}]$ forms a series of grain boundaries. Boundary atom configurations with in-plane offsets of $0[\mathbf{001}]$, $\frac{1}{2}[\mathbf{001}]$, $\frac{2}{2}[\mathbf{001}]$, $\frac{3}{2}[\mathbf{001}]$ are visualized in Fig. 3(a) to Fig. 3(f). Microscopically, these boundaries can be regarded as the chains of coherent locally mirrored twin cores connected by the (100) twin boundaries. Macroscopically, these chains demonstrate contrasts of CAGBs following the $(h-1'0'2)$ plane pairing with the $(h+1'0\bar{'2})$ plane due to microscopic $\frac{h}{2}[001]$ in-plane offsets, as shown in Fig. 3(g). To further clarify the origin of the coherence of these CAGBs, comparative analysis on the low-symmetry $ZrO_2$ and $VO_2$ crystals are presented in Supplemental Material 3 [18]. The similarity with and the distinction from the other two monoclinic oxides emphasize the importance of the special $|[100]| \times \cos(\beta) \approx -\frac{1}{2}|[001]|$ geometric relation of $\beta$-$Ga_2O_3$ conventional cell in the emergence of such highly coherent grain boundary series.

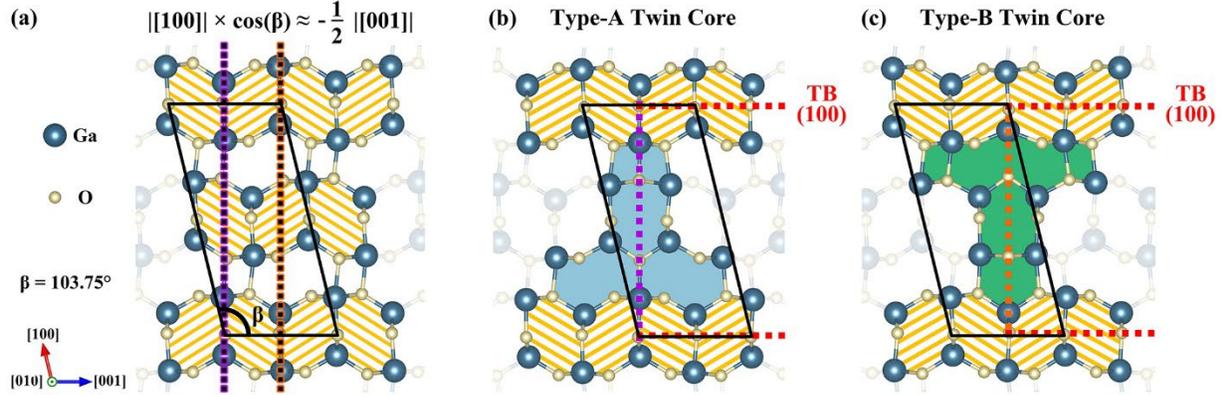

FIG. 2. (a) Schematic illustration of the special geometrical structure of the conventional cell of $\beta$-$Ga_2O_3$ and the in-plane offsets of the local pseudo-mirror-symmetric atom layers. The two distinct long-range, pseudo-coherent atom arrays that are almost perpendicular to the (100) plane are marked by the dashed lines respectively. (b) and (c) Schematic illustration of the coherent locally mirrored twin cores constructed by mirroring across two different coherent atom arrays within one $\beta$-$Ga_2O_3$ conventional cell, which are drawn in (b) purple and (c) orange, respectively. Due to the fact that these two atom arrays are almost perpendicular to the (100) plane, mirroring across these two arrays are geometrically consistent with the (100) twin boundary drawn in red.

Density functional theory calculations are subsequently performed to assess their normalized boundary energy densities ($\sigma_n$). The $(002)/(20\bar{2})-$A CAGB in Fig. 3(b) was found to have the lowest $\sigma_n$ of 0.36 J/m$^2$. Taken together with the lower $\sigma_n$ of the $(202)/(40\bar{2})-$A CAGB in Fig. 3(e) compared to that of the $(202)/(40\bar{2})-$B CAGB in Fig. 3(f), it can be inferred that the type-A twin core is the energetically preferred structure exhibiting higher coherence and lower formation energy density. As the in-plane offsets of the twin cores along the $[001]$ orientation increase (i.e., as the $h$ value increases), the lengths of the (100) twin boundaries connecting them extend, resulting in enlarged $\sigma_n$. Consequently, the energy densities of the CAGBs are expected to increase with higher $h$ values, making the $(002)/(20\bar{2})-$A CAGB the thermodynamically most stable CAGB configuration.

Another important criterion for the emergence of these CAGBs is the surface energy densities of their corresponding crystal planes, as they should be exposable during crystal growth to form a grain boundary. Additional calculations on the surface energy densities of the corresponding crystal planes of the CAGBs in Fig. 3(g) are provided in Fig. 3(h). Evidently, the relatively low surface energy densities of the (002) and $(40\bar{2})$ planes further allow them to become exposed facets during the crystal growth process, which are consistent with previous studies [8]. As the $h$ value increases, the crystal surfaces exhibit more dangling bonds, leading to higher surface energy densities. Taken together, the $(002)/(20\bar{2})-$A CAGB, with its relatively low boundary energy density and the low surface energy density of the (002) plane, is expected to be the most probable CAGB configuration to emerge during crystal growth.

The aforementioned analyses are further verified by growing a 2-inch Fe-doped $\beta$-$Ga_2O_3$ single crystal using the Czochralski growth method under quasi-thermodynamic equilibrium conditions. During this process, we intentionally introduced two (100) twin boundaries and terminated one to form a (100) twin line, as displayed in Fig. 4(a). An optical microscope observation with enhanced per channel contrast of the end boundary of the (100) twin line is presented in Fig. 4(b). The angle between the end boundary and the (100) twin boundary is 103.75°, consistent with the geometry of the $(002)/(20\bar{2})$ CAGB. The total length of the CAGB is as long as 10 μm, indicating its high thermodynamic stability. TEM observation in Fig. 4(c) reveals the microstructure of the CAGB and the (100) twin boundary. Upon the termination of the (100) twin boundary, a $(002)/(20\bar{2})$ CAGB forming an obtuse angle to the (100) twin boundary was immediately generated. Subsequent growth condition interferences redirected the $(002)/(20\bar{2})$ CAGB into a mirrored direction. Notably, the imperfection of the local pseudo-mirror symmetry and the quasi-matching lattice geometry of β-Ga$_2$O$_3$ conventional cell cause lattice strain to accumulate during the growth of the CAGB. The release of this strain mainly involves the formation of dislocations on the CAGB at approximately 80 nm intervals, as showed by the STEM observation in Fig. 4(d). The identification of these dislocations, along with cross-sectional TEM observations of this CAGB and additional observations of another similar CAGB, are presented in Supplemental Material 4 to 6 [18].

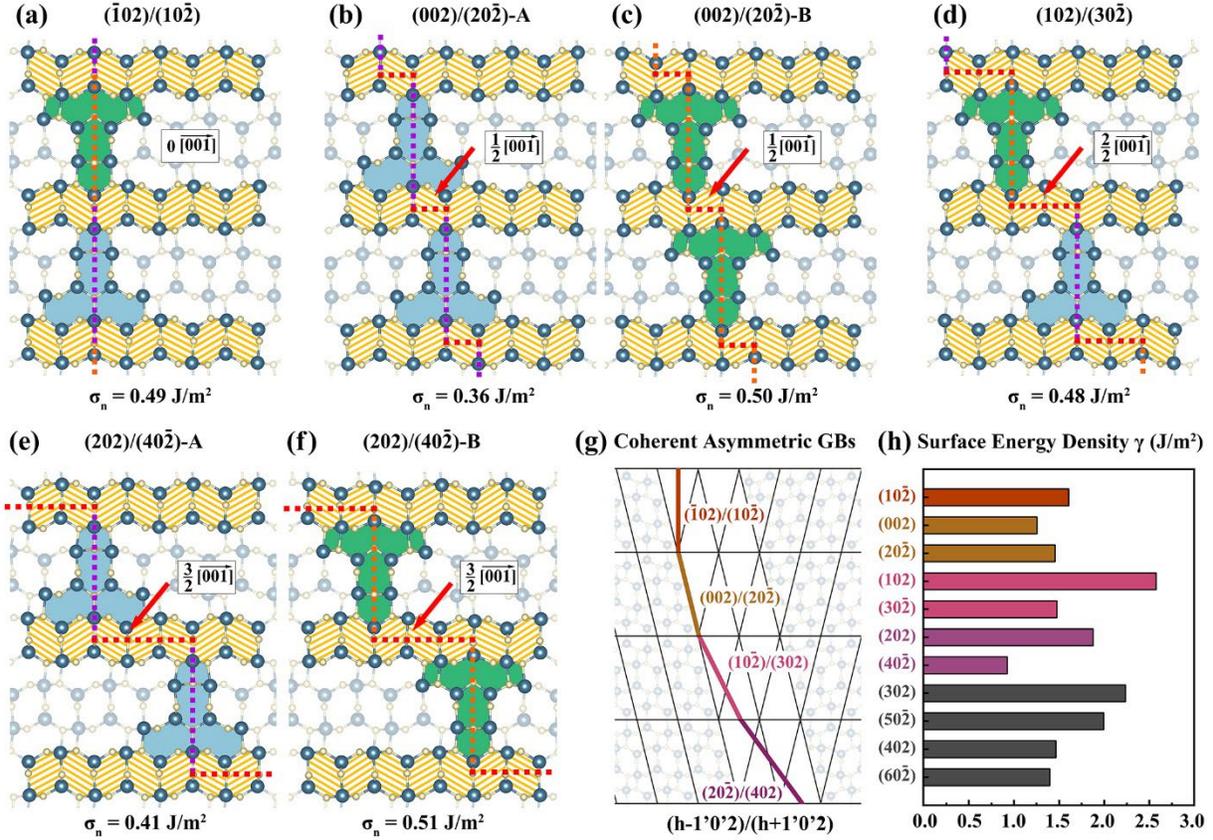

FIG. 3. (a)-(f) Schematic illustrations of the boundaries stacked by the locally mirrored twin cores with different in-plane offsets along the [001] orientation. All coordinates are arranged according to Fig. 1(a) (g) Macroscopic views of the boundaries from (a) to (f), exhibiting CAGB contrast following the $(h-1'0'2)$ plane pairing with the $(h+1'0'\bar{2})$ plane. (f) Surface energy densities of the corresponding crystal planes of the CAGBs shown in (g).

Further HAADF-STEM observations of the atomic-scale structure of this CAGB is carried out. Since the gallium atoms exhibit only a minor difference between the $(002)/(20\bar{2}) - A$ and $(002)/(20\bar{2}) - B$ CAGB, distinguishing the actual atom structure solely through HAADF-STEM is difficult. However, the repetitive mismatched gallium atoms between the $(002)/(20\bar{2}) - B$ CAGB and the observations, as marked by red arrows in Fig. S7 [18], along with the lower $σ_n$ of the $(002)/(20\bar{2}) - A$ CAGB, lead us to conclude that this grain boundary is the $(002)/(20\bar{2}) - A$ CAGB, as presented in Fig. 4(e). The CAGB plays a critical role in the epitaxy growth of β-Ga$_2$O$_3$. Taking epitaxy growth on (100) plane as a demonstration, preparing an epi-ready (100) β-Ga$_2$O$_3$ substrate for step-flow growth mainly involves an off-cut towards either the [001] or [00$\bar{1}$] orientations, followed by high-temperature annealing to form uniform surface steps [5,8,15-17]. As shown in Fig. 5(a), such process exposes the corresponding surface steps with (001) or (20$\bar{1}$) facets as step edge [8]. Due to the symmetry of the β-Ga$_2$O$_3$, a (20$\bar{1}$) twinned defect facet can be formed in the front of the (001) faceted steps, as shown in Fig. 5(b). Three new boundaries, including a free (20$\bar{1}$) surface, a (100) twin boundary, and a $(002)/(20\bar{2}) - A$ CAGB are actually formed to replace the original free (100) and (001) surfaces. The relative stability between the ideal (001) facet and the $(20\bar{1})$ twinned defect facet in Fig. 5(a) and Fig. 5(b) could be assessed through the normalized total energy densities ($\sigma_{ns}$) as follows:

$$\sigma_{ns} = \sum_{i=1}^{j} \sigma_i \cdot S_i, \quad (1)$$

where j is the total number of the free surfaces or the twin boundaries involved, $\sigma_i$ is the energy density of boundary $i$, and $S_i$ is the relative surface/boundary area normalized by the area of the (100) twin boundary shown in Fig. 5(b). Taken together with the additional 0.60 J/m² surface energy density of the (100) plane, the $\sigma_{ns}$ of the ideal (001) facet in Fig. 5(a) is 3.27 J/m². However, based on the 0.92 J/m² surface energy density of the $(20\bar{1})$ plane, the 0.008 J/m² energy density of the (100) twin boundary, and the 0.36 J/m² $\sigma_n$ of the $(002)/(20\bar{2}) - A$ CAGB, the $\sigma_{ns}$ of the $(20\bar{1})$ twinned defect facet in Fig. 5(b) is 3.11 J/m², which is lower than that of the ideal (001) facet. This indicates the twinned defect facet's higher thermodynamic stability and consequently its spontaneous formation during epitaxy growth. The $(20\bar{1})$ faceted steps marked yellow in Fig. 5(a), on the other hand, has no twinned structure since no mirrored free crystal plane have a lower surface energy density than the $(20\bar{1})$ plane, preventing the formation of the twinned defect facets. Taken together, $\beta$-Ga₂O₃ substrates exhibiting surface steps of obtuse (001) and (100) planes shown in Fig. 5(a) are probably not suitable for high-quality epitaxy growth.

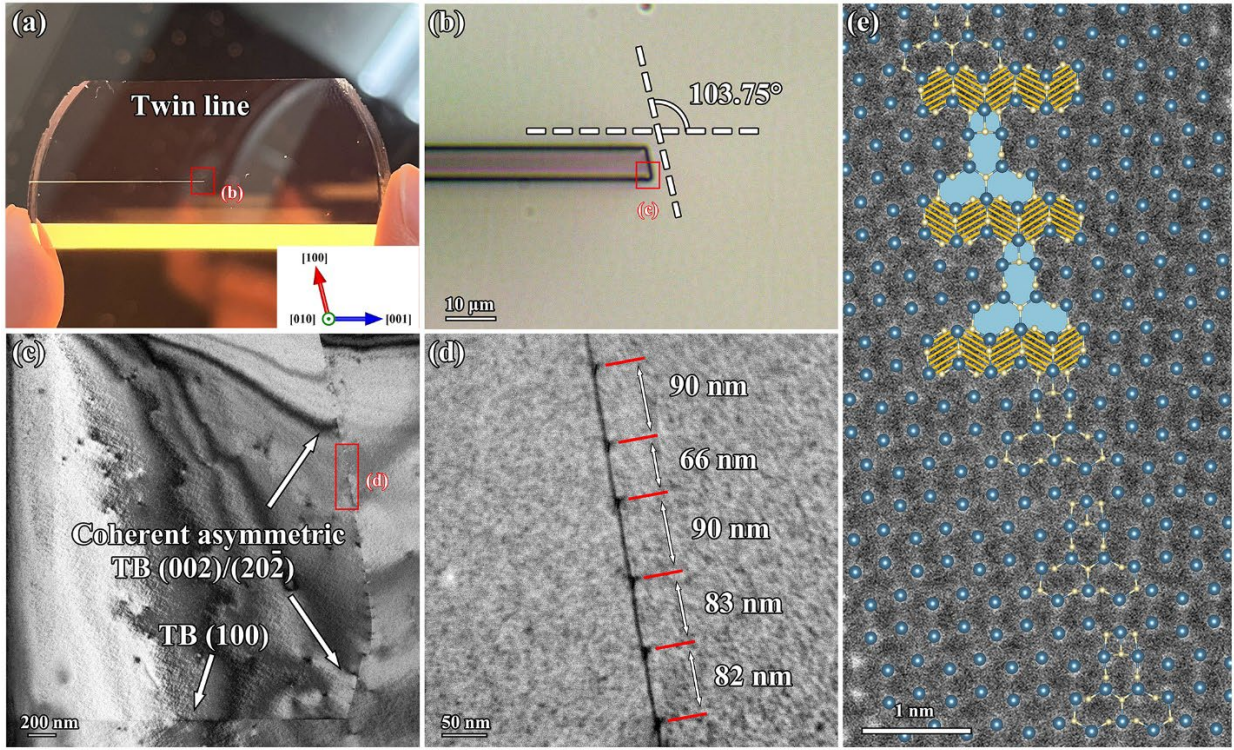

FIG. 4. (a) Top view of the 2-inch substrate containing a full (100) twin boundary and a terminated (100) twin boundary under polarized light. (b) Optical microscopy observation of the end boundary of the terminated (100) twin boundary, showing a macroscopic $(002)/(20\bar{2})$ CAGB geometry. (c) TEM observation of the CAGB. There are two boundaries showing the geometry of $(002)/(20\bar{2})$ CAGB that are symmetric to each other across the (100) plane. (d) STEM observation of the $(002)/(20\bar{2})$ CAGB. (e) HAADF-STEM observation of the CAGB covered with atom structures of the $(002)/(20\bar{2}) - A$ CAGB, which are in good match. All coordinates are arranged according to (a).

Following the above arguments, the necessary and sufficient geometric condition for preventing the twinned defect facet involving low-energy CAGBs is that the substrate surface planes should not expose the obtuse dihedral formed by the (001) and (100) planes. Such planes lie within the obtuse dihedral angle between the (100) and (001) planes, as indicated by the green arc shown in Fig. 5(c). On the contrary, the orange arc in Fig. 5(c) consisting of crystal plane forming triangles with obtuse dihedral angle of the (100) and (001) planes, such as the (101) and (201) planes, should be avoided when choosing substrates. A previous study has shown that an off-cut of the (100) plane substrate towards the $[00\bar{1}]$ orientation, which exposes the $(20\bar{1})$ facet, is superior to an off-cut towards [001] orientation that exposes the obtuse (001) and (100) surface steps [8]. This confirms the aforementioned selection rule. Furthermore, for high-quality epitaxy on the (001) substrates, an off-cut towards [$\bar{1}$00] orientation should be superior to an off-cut towards [100] orientation, since the former exposes ($20\bar{1}$) facets while the later exposes obtuse dihedral formed by (001) and (001) facets.

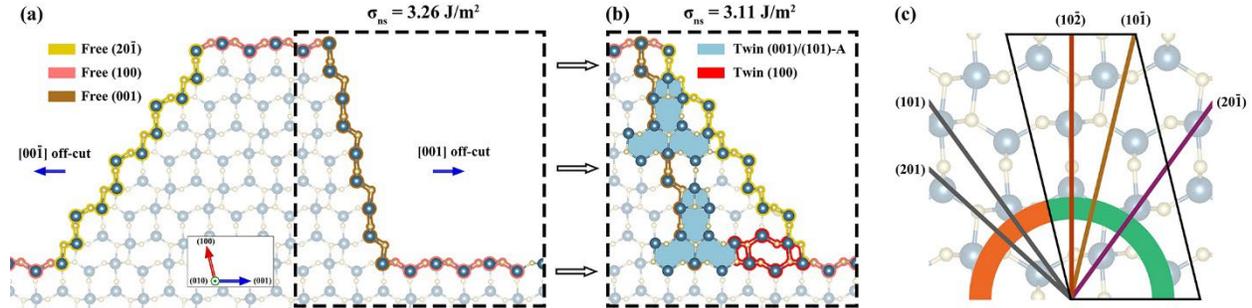

**FIG. 5.** (a) Edge of the surface step on the (100) substrate of β-Ga$_2$O$_3$ exposing (001) facet either ($20\bar{1}$) or (001) facets. (b) Spontaneously formation of a twinned nucleus at the (001) facet with a lower $\sigma_{ns}$. (c) Substrates within the orange region is expected to exhibit (001) facet in (a) while substrates within the green region could avoid it. Specifically, the $h \cdot k$ of a ($hkl$) crystal should be negative.

In summary, we comprehensively investigated the highly coherent (100) twin boundary and a series of coherent asymmetric grain boundaries (CAGBs) in the low-symmetry β-Ga$_2$O$_3$ crystal. Through the analyses of the local pseudo-mirror-symmetric atomic layers and the special geometric relation of the β-Ga$_2$O$_3$ conventional cell, we show that these boundaries exhibit high coherence under the coincident-site lattice (CSL) model, with 4 out of 5 atoms place in pseudo-coincident sites. With DFT calculations, Czochralski growth experiment, and atomic-scale characterizations, we confirmed the stability of the $(002)/(20\bar{2}) - A$ CAGB configuration, whose formation energy density $\sigma_n$ is 0.36 J/m$^2$. This low $\sigma_n$ should primarily be attributed to the long-range, pseudo-coherent atom arrays in β-Ga$_2$O$_3$ that are quasi-perpendicular to the (100) plane. This relatively low $\sigma_n$, along with the ultralow formation energy density of the (100) twin boundary, leads to the spontaneous formation of a twinned defect facet at the obtuse dihedral of the (100) and (001) planes during step-flow epitaxy growth. A consequent selection rule of β-Ga$_2$O$_3$ substrate orientation is derived to prevent its formation and endow high-quality epitaxy. Based on this study, we provide perspectives into the physical origins of the grain boundaries in the low-symmetry β-Ga$_2$O$_3$ system, while emphasizing the importance of analyzing the atomic-scale pseudo-symmetric structures when studying grain boundaries in low-symmetry crystal systems.

*Acknowledgments*—This work was supported by the 'Pioneer' and 'Leading Goose' R&D Program of Zhejiang (2023C01193), the National Natural Science Foundation of China (52202150, 22205203, 62204218), Fundamental Research Funds for the Central Universities (226-2022-00200 and 226-2022-00250), the Foundation for Innovative Research Groups of the National Natural Science Foundation of China (61721005), the National Postdoctoral Program for Innovative Talents (BX20220264), the National Program for Support of Topnotch Young Professionals, Leading Innovative and Entrepreneur Team Introduction Program of Hangzhou (TD2022012), and the Open Fund of the State Key Laboratory of Optoelectronic Materials and Technologies (Sun Yat-sen University).